\documentclass[a4paper,11pt]{article}

\usepackage{blois}
\usepackage[pdftex]{graphicx}

\title{ULTRA-HIGH ENERGY COSMIC RAYS FROM RADIO GALAXIES REVISITED}
\author{J\"ORG P. RACHEN}
\address{Max-Planck-Institute for Astrophysics, 85741 Garching, Germany}

\begin{document}
\vspace*{4cm}
\maketitle\abstracts{
A striking concentration of ultra-high energy cosmic ray (UHECR) events observed by the Pierre Auger Observatory around the direction of the nearby radio galaxy Centaurus A revives the idea that radio galaxies may be dominant sources of UHECR. In this paper, we give a brief overview about processes which may accelerate protons and nuclei in radio galaxies, and their relation to jet power, radio morphology and cosmic source density. We argue that, except for the most powerful FR-II radio galaxies, processes in radio lobes are unlikely to explain the origin of UHECR. However, Fermi acceleration of protons at internal shocks in the ``blazar-zone'' of \textit{all} radio galaxies, and their photohadronic conversion into neutrons, may lead to the ejection of ``UHECR-beams'', which remain collimated over several Mpc. Consequences of this hypothesis for the interpretation of the UHECR event distribution, in particular for the special case of Centaurus A, are discussed.}

\section{Fermi acceleration in radio galaxies}

First order Fermi acceleration is considered to be the most promising physical process for the acceleration of cosmic rays, because it generates an exponential grow in the energy of particles magnetically confined in the vicinity of strong shock waves. As shock waves are quite frequent in the universe, shock accelerators might be an ubiquitous phenomenon, but for the highest observed energies, targeted by the Pierre Auger Observatory, objects harboring the right physical conditions become sparse. Only jets of radio galaxies\,\cite{RB93,MPR01} and gamma-ray bursts\,\cite{GRBrefs} have been shown to be able to provide the energy and flux observed for the highest energy cosmic rays. Also larger scale shocks have been discussed as sources, but these generally require special assumptions for the acceleration process.\cite{KRB97} 

The maximum energy attainable in a shock accelerator, which we allow here to move with bulk relativistic speed, can simply be estimated as
\begin{equation}\label{Emax}
E_{\rm max} \sim Z e B' R' \beta_{\rm sh}' \Gamma
\end{equation} 
where $Ze$ is the particle charge, $R'$ and $B'$ are the comoving size and magnetic field of the acceleration region, $\beta_{\rm sh}'$ the velocity of a shock wave traveling through the comoving region, and $\Gamma$ the bulk Lorentz factor of the entire acceleration region in the observers frame. This differs from the simple confinement limit ($E'{<} Z e B'R'$) by the factor $\beta_{\rm sh}'$, which arises from considerations on acceleration time scale (${\propto} 1/{\beta_{\rm sh}'}^2$) and shock life time (${\propto} 1/\beta_{\rm sh}'$). Note that all these factors are crude estimates, assuming isotropic geometries and diffusion coefficients, so that this limit should be considered valid only by order of magnitude. 

In radio galaxies shock acceleration of electrons is known to take place where jets interact with the surrounding lobe, as this is the only consistent explanation for the origin of synchrotron radiation observed from such structures. The most prominent features of this kind are the so-called ``hot spots'' in the most powerful radio galaxies, which usually outshine the central region of the AGN in radio frequencies. According to a classification scheme introduced by Fanaroff and Riley,\cite{FR74} these powerful sources are known as FR-II radio galaxies. With $B\sim 200{-}500\,\mu$G, $R\sim 0.3{-}3\,$kpc, and $\beta_{\rm sh}\sim 0.2{-}0.5$ for observed FR-II hot spots,\cite{M99} we see from Eq.(\ref{Emax}) that maximum proton energies ${>}100\,$EeV can be produced, and it was shown that this contribution can explain the observed UHECR flux and spectrum.\cite{RB93} If the large scale jets of FR-II galaxies contain a considerable amount of heavy nuclei, these may be accelerated by the hot spot shocks to even higher energies.

The disadvantage of this model is that FR-II radio galaxies are very rare. The nearest one is located at a distance of about $100\,$Mpc, and both the smooth GZK cutoff in the observed spectrum and the arrival direction distribution argue against FR-II radio galaxies as the dominant source of UHECR. If we try to save the day for radio galaxies by na\"ively extending the model to the weaker FR-I galaxies, we run into deep trouble: First, the shocks at the end point of FR-I jets are much less powerful and slower than in the FR-II case, and the magnetic fields are lower. For the inner lobes of Centaurus A, $B\sim 10\,\mu$G, $R\sim 1\,$kpc and $\beta_{\rm sh}{\sim}0.1$ miss the goal of $100\,$EeV by almost two orders of magnitude.\cite{CenA_inner} The situation is even worse inside the giant lobes, which surround both FR-I and FR-II galaxies: Although the confinement condition $E{<}ZeBR$ easily holds in them even for $Z{=}1$, with typical $B{\sim}3\,\mu$G and $R{\sim}300\,$kpc, plasma flows in giant lobes are likely to be subsonic, thus shocks are weak or do not exist at all.\cite{Kron} All in all, UHECR production at shocks in the lobes of radio galaxies or their substructures seems disfavored. 

\section{Neutron beams from blazars}

Another possible acceleration site in radio-loud AGN has been pointed out by Mannheim in 1993:\,\cite{M93} If the compact, relativistic ($\Gamma\sim 10$) AGN jets, now often referred to as the ``blazar zone'' of the AGN, contain a considerable fraction of hadronic matter, then internal shocks, commonly assumed to produce the strong gamma-ray flares in blazars, would be able to accelerate protons to UHECR energies. Charged particles in these compact zones would be magnetically confined and lose their energy by adiabatic expansion of the jet. However, in interactions with ambient photons, protons produce three types of neutral particles, which can leave the AGN jet unattenuated in relativistically boosted, collimated beams: photons, neutrinos, and neutrons. The effects of escaping photons and neutrinos have been extensively discussed in the literature,\cite{hadronic} while the contribution of neutron beams has only been considered to place upper limits on the possible neutrino or gamma ray emission from this process.\cite{MPR01} 

In the energy range of interest here (${>}60\,$EeV) the average decay length of relativistic neutrons exceeds $600\,$kpc. Therefore, neutron beams from a blazar, boosted with $\Gamma{\sim}10$ into an angle of a few degrees, would remain collimated until they have left almost all magnetized structures around their production site. [An exception may be radio galaxies in the center of galaxy clusters, like M87.] At this point, most neutrons have decayed into protons, but the resulting ``UHECR beam'' would still remain collimated as intergalactic magnetic fields are too weak (${<}1\,$nG) to produce significant deflections. Only occasional encounters with magnetized wind zones around galaxies would slowly dissipate the proton jet on a time scale comparable to that of the GZK process. Of course, as neutrons decay with constant probability, a fraction of the produced UHECR, comparable to the fraction of the path length of the neutron beam through the giant lobe to the decay length of the neutron, would always be caught and isotropized by the lobe of the source. This would produce an isotropic ``UHECR glow'' of radio galaxy lobes, which could not contain any heavy nuclei as these cannot escape from the blazar zone. 

In order to compare the relevance of this process to the acceleration in lobes, in particular in the hot spots in FR-II radio galaxies, we return to Eq.(\ref{Emax}). Unfortunately, the physical parameters of the blazar zone are not directly observable and are quite model dependent. However, noting that the magnetic energy flux of the jet is $\phi^B_{\rm jet} \sim (B^2/8\pi)(\pi R^2 c)\Gamma^2$, and the total jet flux under minimal energy conditions is $\phi_{\rm jet} \sim 2\phi^B_{\rm jet}$, we can easily transform Eq.(\ref{Emax}) into
\begin{equation}\label{power}
E_{\rm max} \sim Z \beta_{\rm sh}' \big[4 \alpha \hbar \phi_{\rm jet}\big]^{1/2} 
\end{equation} 
with $\alpha$ being the fine structure constant. Requiring again $E{\sim}100\,$EeV, this time strictly for $Z{=}1$, and assuming $\beta_{\rm sh}'{\approx}1$ as typical for internal shocks in blazar zones, we obtain $\phi_{\rm jet}{\sim}10^{45}\,$erg/s. Intriguingly, this is again beyond the demarcation line between FR-I and FR-II radio galaxies, even if we assume a proton-to-electron ratio of $k_p{\approx}100$ in the jet. [Typical jet power estimates arise from minimal energy considerations, assuming only relativistic electrons and magnetic fields in the jet. The presence of relativistic protons scales this with $(1+k_p)^{4/7}$]\,\cite{minimumenergy}. The situation is still favorable compared to radio lobes, as the breakdown of morphological structures in the lobes drops $E_{\rm max}$ by orders of magnitude when going from FR-II to FR-I radio galaxies, while in Eq.(\ref{power}) $E_{\rm max}$ decreases only gradually with the square root of the jet power. Thus, a significant fraction of blazars in FR-I radio galaxies can contribute to the UHECR above $30\,$EeV, but for the highest energy events we are left in the same dilemma: only the powerful, but rare FR-II radio galaxies can produce them. 

\section{Centaurus A: a radio zombie?}

\begin{figure}
\parbox{0.4205\textwidth}%
{\caption{Sketch of the giant radio lobe of Centaurus~A, together with five nearby Auger events above $57\,$EeV. Around each Auger event, a one degree (filled), three degree (thick dotted) and five degree (thin dotted) circle is shown, corresponding to the direction reconstruction error of the Pierre Auger Observatory and the likely range of deflection angles in cosmic magnetic fields. The energy of the events is shown in their inner circle, given in EeV. For the Centaurus~A lobe, regions of more intense radio emission are indicated (concentric circles), as well as the central region (hatched circle) harboring the entire inner structure (i.e. the galaxy, the jet and the inner lobes). Also shown are the direction of the active jet (thin solid lines), and the putative direction of the UHECR jet associated with the giant lobe production (thick dashed line, see text for explanation). Note that all five Auger events are aligned with the axis of this putative jet. There are no other Auger events above $57\,$EeV in the area shown here. The figure might show slight geometric distortions compared to reality, as a ${\sim}30^\circ{\times}20^\circ$ patch of the sphere has been projected on a rectangular grid.}\label{CenA}}%
\hfill\parbox{0.57\textwidth}{\centerline{\includegraphics[height=10.5cm]{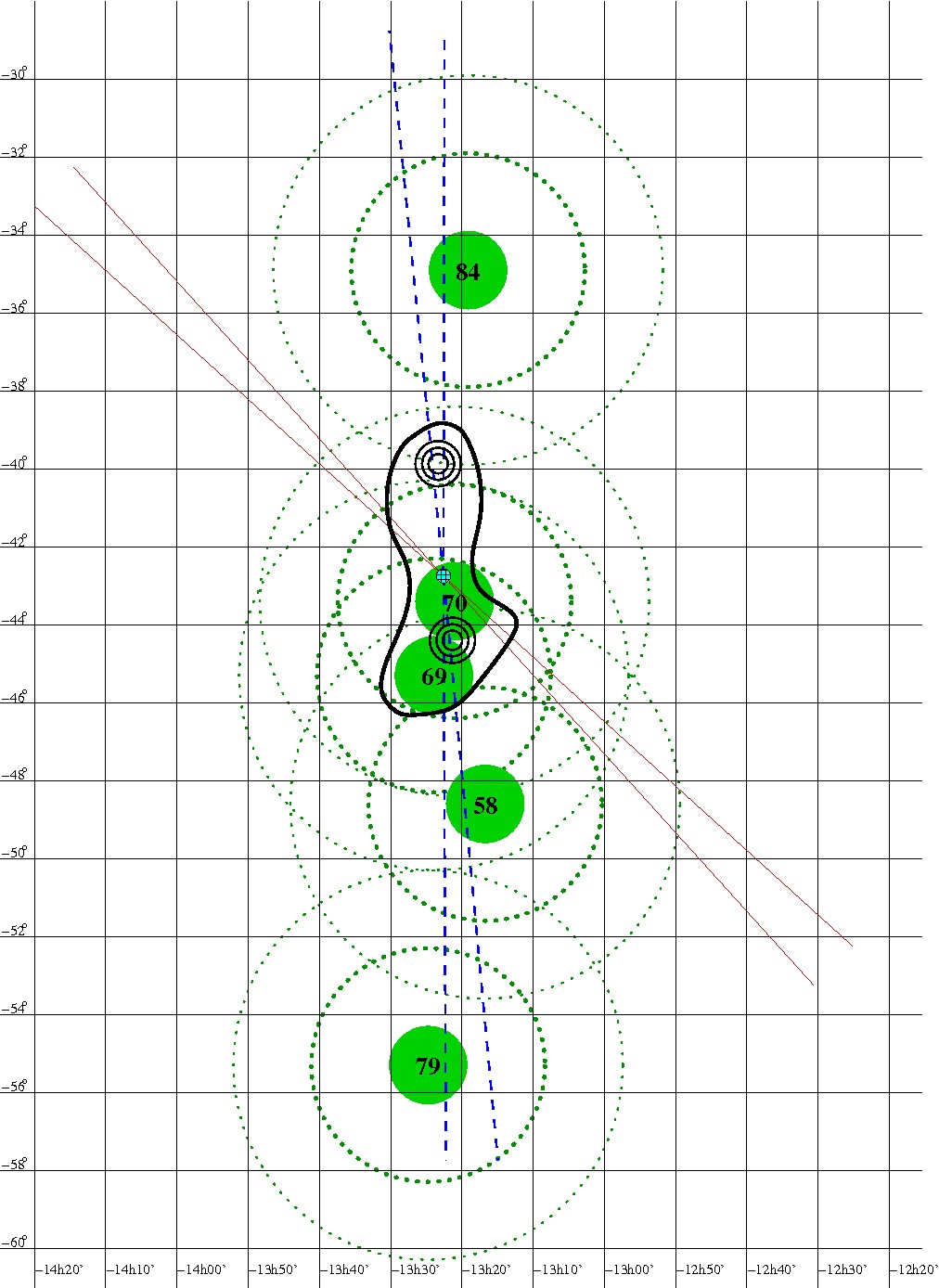}}}
\end{figure}

Although not discussed explicitly in the Auger papers,\cite{Auger} the strong concentration of events (10 of 27) in a ${\approx}15^\circ$ circle around Centaurus~A (Cen~A), the nearest radio galaxy, stroke almost every astronomer who has looked at the map so far. Indeed, taking the large extended lobes of Cen~A as a target, and allowing for up to $3^\circ$ deflection in galactic and extragalactic magnetic fields, there are 3 events with directions consistent with the position of Cen~A. Assuming a Poisson distribution of the 27 events, the estimated chance probability for this association is $P \sim 2{\times}10^{-4}$. The arrival directions of two events are enclosed in the lobe ($P \sim 2{\times}10^{-4}$), and extending the allowed deflection to $5^\circ$ increases the number of consistent events to four ($P \sim 6{\times}10^{-5}$). A fifth event is at an angular distance of ${\approx}10^\circ$ from the center of Cen~A (see Fig.~\ref{CenA}). We are therefore led to ask whether Cen~A could indeed be the source of these events.

Looking at the physics, the answer is discouraging. Cen~A is a FR-I radio galaxy, not even a particularly strong one. From the discussion above, we may conclude that none of its extended radio structures, neither the inner lobes of about 1 kpc, nor the giant lobe of several 100 kpc extension, is capable to accelerate these particles. Considering possible UHECR beams from the blazar zone, the situation doesn't look better: with an estimated\,\cite{CenA_power} $10^{42}$ to $10^{43}\,$erg/s, the jet power fails the requirements for accelerating particles to the observed energies by two orders of magnitude. Thus, if the association of arrival directions with Cen A is not accidental, we seem to need some voodoo to explain it.

The required magic may be found in the nature of radio galaxies. Their complex shapes and spectral features suggest that they go through temporary activity phases, lasting not significantly longer than $10^8\,$years, in which the sources may appear as strong FR-II radio galaxies, while at other times they appear as weaker FR-I radio galaxies. The lobes created during their activity phases keep up their pressure carried by magnetic fields and relativistic particles for a long time. Eventually, they detach from their original source, driven by buoyancy in a sourrounding medium in a galaxy group or cluster, becoming ``radio ghosts''.\cite{ghosts} Such radio ghosts are observed as cavities in the X-ray emitting medium of dense galaxy clusters,\cite{ghosts_observed} and, due to their highly ordered magnetic fields, they can serve as excellent scattering centers for proton beams, isotropizing them up to the highest cosmic ray energies.\cite{ghosts}

Cen~A is a tiny active radio galaxy, spanning with its inner lobes only less than $10\,$kpc, but it is surrounded by a giant lobe with a length of ${\sim}600\,$kpc, which is rather typical for the most powerful FR-II galaxies. Moreover, the active jet is inclined by about $45^\circ$ to the main axis of the giant lobe. It appears improbable that the present jet has produced the giant lobe, rather it seems likely that the latter is a remnant of a more active phase of Cen~A. In the spiritualist terminology, Cen~A can be seen as a resurrected, weak radio galaxy living in the dead body of a former life --- a ``radio zombie''.

The age of the giant lobe can be estimated to be less than $3{\times}10^7$ years,\cite{CenA_GL} comparable to the time scale of a confined $100\,$EeV proton to escape in a random walk. The powerful jet which produced this lobe could have sent an UHECR beam in the direction of its main extension. A large part of this beam would have been isotropized in the giant lobe, but some part would have penetrated it, lighting up surrounding radio ghosts from earlier activity phases of Cen~A. Indeed, all five Auger events shown in Fig.~\ref{CenA} are perfectly aligned with the main lobe axis, in a distance form a core less than the decollimation length of an UHECR jet. The chance probability for this alignment is difficult to estimate --- just asking for the Poisson probability of finding 5 of 27 events in a $3^\circ{\times}30^\circ$ stripe gives $P\sim 2{\times}10^{-8}$.

\section{Consequences and outlook}

With our current knowledge on radio galaxies, and the current statistics of UHECR events, all of the above is pure speculation. More work, both theoretical and observational, needs to be done in order to confirm or falsify the idea that UHECR are emitted as collimated neutron beams from compact jets in radio galaxies. 

First, the theory of neutron beams emitted from blazars needs to be detailed (Rachen and En{\ss}lin, in preparation). Then, we need a significantly improved statistics of UHECR events, to show whether the correlation with Cen~A in general, and the alignment with its lobe main axis in particular, is significant. The Pierre Auger Observatory may deliver the required statistics in a few years of operation. But even if this correlation would be confirmed, we would still need to establish the existence of radio ghosts in the directions of the aligned UHECR events outside the giant lobe. Low frequency radio observations, soon to be realized with LOFAR,\cite{LOFAR} may be the key to this, but to observe Cen~A, a similar facility would be needed in the southern hemisphere.

Second, the model would imply that the highest energy cosmic rays are dominated by protons. However, this prediction should not be overinterpreted, because a detailed comparison of this process to other source scenarios, like hot-spots in FR-II galaxies and gamma-ray bursts, and their composition predictions need to be done. It would be quite suprising if just one class of sources produced all UHECR. 

Third, as already mentioned, production of UHECR in blazars implies significant cosmic neutrino fluxes, peaking at about $1\,$EeV. Mannheim, Protheroe and Rachen\,\cite{MPR01} derived an upper limit for these neutrino fluxes, but to allow experimental tests of the present model we would need a lower limit, i.e., the neutrino flux necessary for the model to work. In a few years, IceCube will reach comparable limits,\cite{IceCube} either confirming the model by observation of cosmic neutrinos, or putting severe constraints on it.

Finally, another corollary of the present model is that gamma ray emission of blazars is of hadronic origin, rather than being produced by inverse Compton scattering. The latter model is widely accepted in the gamma ray community, and although hadronic blazar models have never been ruled out by data, they are hardly ever considered. A dedicated, phenomenological research on how to distinguish the hadronic model from the inverse Compton scenario by suitable observational campaigns would be needed. It must be clear that, due to completely incompatible physical conditions assumed in these models, not both can be true: blazar jets as sources of UHECR, and gamma-ray emission of blazars dominated by inverse Compton scattering off electrons. Multifrequency observations of blazars and their theoretical interpretation remain therefore crucial for our understanding of the origin of the highest energy cosmic rays.

\begin{footnotesize}
\section*{Acknowledgements}

I wish to thank Eli Waxman and Michael Kachelrie{\ss} for pointing out the connection between jet power and maximum proton energy to me, Hagar Landsman for enlightening discussions on the status and prospects of IceCube, and the organizers of the \textit{Rencontres de Blois} for arranging this great conference and their tolerance to accept my overdue contribution. I~am indebted to Torsten En{\ss}lin for his support on this and related research, and for his comments on the manuscript.

\end{footnotesize}
\end{document}